\documentclass[useAMS,usenatbib]{mn2e}

\usepackage{graphicx}
\usepackage{multirow} 
\def\lesssim{\la}
\def\gtrsim{\ga}

\title[]{Origin Scenarios for the Kepler 36 Planetary System}
\author[Quillen, Bodman \& Moore]{Alice C. Quillen, Eva Bodman, \& Alexander Moore
\\
Department of Physics and Astronomy, University of Rochester, Rochester, NY 14627, USA \\
}

\begin{document}
\maketitle
\begin{abstract}
We explore scenarios for the origin of two different density 
planets in the Kepler 36 system in adjacent orbits
near the 7:6  mean motion resonance.
We find that
fine tuning is required in the stochastic forcing amplitude, the migration rate
and planet eccentricities to allow two convergently migrating planets to bypass mean motion resonances such 
as the 4:3, 5:4 and 6:5, and yet allow capture into the 7:6 resonance.  
Stochastic forcing can eject the system from resonance
causing a collision between the planets, unless the disk causing migration and the stochastic forcing is depleted soon after resonance capture.

We explore a scenario with approximately
Mars mass embryos originating exterior to the two planets and migrating inwards
toward two planets.  We find that
gravitational interactions with embryos can nudge the system out of resonances. 
Numerical integrations with about a half dozen embryos can leave the two planets
in the 7:6 resonance.  Collisions between planets and embryos have a wide
distribution of impact angles and velocities ranging from accretionary to disruptive.
We find that impacts can occur at sufficiently high impact angle and velocity that
  the envelope of a planet could have been stripped, leaving behind a dense core.
Some of our integrations show  the two planets exchanging locations, allowing the outer planet
that had experienced multiple collisions with embryos to become the innermost planet.
A scenario involving gravitational interactions and collisions with embryos may account for
both the proximity of the Kepler 36 planets and their large density contrast. 

\end{abstract}

\section{Introduction}

The Kepler mission \citep{borucki10} has detected over 2300 planet candidates \citep{batalha13} and 
about one third of KeplerÕs candidates are associated with compact multiple transiting systems \citep{lissauer11}.
A statistical analysis, focused on the probability that binary stars are the most likely contaminant, find that most of the multiple planet candidates are real planetary systems \citep{lissauer12} and that approximately 20\% of the total number of planet candidate systems are multiple planet systems \citep{latham11,lissauer11}. 

Among the newly discovered planetary systems is the exotic Kepler 36 system, that hosts 
two transiting planets Kepler-36b and c, with
orbital periods of 13.8 and 16.2 days and masses of $m_b=4.1$ and
$m_c = 7.5 M_\oplus$, respectively \citep{carter12}.
Of all the multiplanet systems, Kepler-36 has the smallest fractional separation between 
any pair of adjacent orbits, and this pair also has one of the largest planet density contrasts  \citep{carter12}.
The  estimated density for Kep36c is only $\rho_c = 0.89$ g~cm$^{-3}$ and that for the less massive
  Kep36b  is high at $\rho_b = 7.46$ g~cm$^{-3}$. 
The two planets' orbits are separated so that they are just exterior to the 7:6 first order mean motion resonance.
Numerical integrations by \citet{carter12}, constrained by transit timing observations, 
not only have measured the planet masses, but
restrict the planet orbital eccentricities to less than 0.04.
Numerical integrations by
\citet{deck12} illustrate that the system is chaotic likely due to the proximity of
the 7:6 resonance and a weak, but high order, two-body resonance.

Because the Kepler systems are compact, each planet fills a sizable fraction of its Hill sphere.
This is illustrated in Table \ref{tab:tab1} showing ratios of planetary radii to semi-major axes
and Hill radii.
The distance between the planets semi-major axes
in units of the outer planet's radius is only $(a_c-a_b)/R_c = 84.9$
and only 4.8 mutual Hill radii.\footnote{The mutual Hill radius 
$r_{mH} \equiv \left({m_b + m_c \over 3 M_*}\right) \left({a_b +a_c \over 2}\right).$}  
As explored by \citet{deck12}, the planets are so close that they 
exhibit chaotic evolution.
Because of the planets' proximity and large size compared to their semi-major axis, 
it is important to check for collisions during numerical integrations while exploring scenarios
for the origin of the system.
Because the planets have short orbital periods, their orbital velocities are high 
at approximately 91 and 86 km/s for Kep 36b and 36c respectively.    The orbital velocities can be compared
to the escape velocities from the planets that are approximately $18.8$ and $16.1$ km/s.  
Collisions between bodies in this system
would be at high velocity compared to the planets' escape velocities.
Scaling laws predicting the outcome of collisions for terrestrial bodies depend on 
the impact velocity in units of the mutual escape velocity, the mass ratio of projectile to target mass,
impact angle and body constituents \citep{asphaug10,marcus10,stewart12,leinhardt12}.
Collisions at velocities above the mutual escape velocity can be highly erosional. 
A scenario involving a high velocity collision, such as proposed for the formation of Mercury,
and accounting for Mercury's high density \citep{benz07}, could account for the high density of Kep 36b.

In this paper we investigate scenarios for the origin of the Kepler 36 system, focusing not only on
the proximity of the two planets but on the possibility of collisions between bodies.

\begin{table}
\vbox to 45mm{\vfil
\caption{\large Some dimensionless quantities in the Kepler 36 planetary system 
     \label{tab:tab1}}
\begin{tabular}{@{}lccc}
\hline
                                                   &  Planet b   & Planet c \\
Planet mass/Stellar mass            &   $1.15\times 10^{-5} $    &  $   2.09  \times 10^{-5} $        \\
Orbital velocity/Escape velocity    &   4.8            &5.3 \\
Semi-major axis/Hill radius           &   63.9          &52.3\\
Hill radius/Planet radius               &    29.0        &16.0  \\
Semi-major Axis/Planet radius     &     1852      &       838\\
\hline
\end{tabular}
{\\ These quantities are computed using masses, planetary radii and orbital periods by \citet{carter12}.
}
\vfil}
\end{table}

\section{Fine tuning in the Stochastic Migration Scenario}

In many multiple planet systems, pairs of planets
appear to be close to mean-motion resonances, that is,
the ratio of their orbital periods are close to the ratio of two small integers.
Because mean motion resonances are narrow (with width scaling with
the ratio of planet to stellar mass),
the existence of near-resonant planet pairs is usually ascribed
to resonant capture due to convergent migration \citep{snellgrove01}
(though see \citealt{petrovich13,owen13} exploring in-situ formation).

Previous works have explored the properties of planet semi-major axis or period distributions that are predicted
from stochastic migration models \citep{rein13}.  It is assumed that the semi-major axes and eccentricities
of the planets can vary to due to torques transferred
between embryos or planets to a gaseous disk via driving of density waves \citep{ward97,tanaka02}.
The gas disk is no longer present, having long since dissipated. 
A two planet system near the 7:6 
was not predicted by the in-situ formation scenario
or the stochastic migration scenario and even high mass planet pairs in the 4:3 resonance
are difficult to explain \citep{rein12}.   We first explore whether 
a stochastically migrating two planet system would be likely to be subsequently 
found near the 7:6 mean motion resonance.   Concurrent to our study is the recent numerical exploration
by \citet{pardekooper13} who have shown that stochastic migration is a viable way to allow
capture into the 7:6 resonance for the Kepler 36 planets.   Our approach in this section 
is primarily analytical rather than focused
on hydrodynamical simulations and so is complimentary to their study.

\subsection{Sensitivity to migration rate and initial planet eccentricity}

We consider a setting where the two planets approach one another due to convergent migration.
As the system drifts it encounters mean motion resonances between the two planets.
A first order $j:j-1$ resonance is a commensurability where $j n_c \approx (j-1) n_b$ where $n_b, n_c$
are the mean motions of the two planets.
Capture into a mean motion resonance is likely if the planet eccentricities are low, the planets approach each other, and the relative drift rate
is smooth and slow (e.g., \citealt{M+D,henrard82,borderies84,quillen06, mustill11}).

Under what conditions can a smoothly migrating system bypass lower $j$ first order mean motion resonances
such as the 3:2 and 4:3 resonances,
but capture in a higher $j$ one such as the 7:6 resonance?
Higher $j$
first order mean motion resonances are stronger, wider and have faster libration frequencies than lower $j$ ones.
If the migration rate is sufficiently high, then first order 
resonances such as the 3:2, 4:3, 5:4 and 6:5 would be bypassed, but
capture into the 7:6 resonance would still occur.

Capture of a low mass particle into a first order mean motion resonance with a planet can be modeled with a one dimensional Hamiltonian system (e.g., \citealt{M+D,henrard82,borderies84,malhotra90,quillen06,mustill11})
\begin{equation}
H(\Gamma,\phi) = A\Gamma^2 + b \Gamma + \epsilon \Gamma^{1/2} \cos \phi, \label{eqn:Ham}
\end{equation}
where $\Gamma \equiv \sqrt{a}(1 - \sqrt{1-e^2})$ is a Poincar\'e momentum for a particle
with semi-major axis $a$ and eccentricity $e$.   
We consider capture of a massless particle
into resonance with a planet.  We neglect the fact that the Kepler 36 system
consists of two planets with similar masses. To order of magnitude, this is justified 
as the more general formulation by 
\citet{mustill11} only differs by factors of order unity from the case of a massless particle near a planet.
Above we have assumed units $GM_*=1$ with $G$ the gravitational constant and $M_*$ the mass
of the host star.  The resonant argument,
\begin{equation} 
\phi = j \lambda - (j-1) \lambda_p + \varpi, \end{equation} 
corresponds to a slowly varying angle in proximity to the $j:j-1$ resonance for an object external to a planet.    Here $\lambda, \lambda_p$ are 
the mean longitudes of particle and planet, respectively, and $\varpi $ is the particle's longitude of pericenter. 

The coefficients, $A,b,\epsilon$, depend upon the integer $j$ and 
the ratio of particle to planet semi-major axes, $\alpha \equiv a/a_p$.    
For a drifting system, the proximity to resonance, described by frequency $b$, is a function of time.
This model is derived by expanding the Keplerian Hamiltonian near resonance and 
adding the lowest order eccentricity term in the disturbing function.
The coefficients
$A \approx- j^2$ and $\epsilon \approx \mu_p f_{27}(\alpha)$ \citep{M+D,quillen06,mustill11} 
where $f_{27}(\alpha)$ is a function
of Laplace coefficients and is given in the appendix by \citet{M+D}.  Here $\mu_p$ is the ratio
of planet to stellar mass.   
Because the 7:6 mean motion resonance has a moderately high $j$ value, we can estimate
the dimensions of the resonance using the asymptotic limit;
$\alpha \to 1$.  In this limit, the Laplace coefficient, 
$b_{1/2}^{(j)} (\alpha) \sim \ln (1-\alpha)$ \citep{quillen11},  and 
the coefficient $f_{27}(\alpha) \sim - (1-\alpha)^{-1} \sim - j$  for $\alpha = {(j-1)/ j}$ set by the resonance condition.

The particle can only be captured into resonance if the drift rate, $\dot b$,  
is slower than a critical rate that can be estimated through dimensional arguments and delineates
the adiabatic limit \citep{quillen06,mustill11}. 
Because the frequency $b \approx j n -(j-1)n_p$ is small in resonance,  
the Hamiltonian contains a single dimensional unit of time 
\begin{equation} 
t_{lib} \sim  |\epsilon|^{-2/3} |A|^{-1/3} \label{eqn:tlib} 
\end{equation} 
that is also approximately the inverse of the libration frequency in resonance.
The Hamiltonian contains a single dimensional unit of momentum  
\begin{equation} 
\Gamma_{res} \sim  \left|{\epsilon \over A}\right|^{2/3}, \label{eqn:gamma_res}
\end{equation}  
that is approximately the size scale of eccentricity oscillations for a system  in resonance at low eccentricity.

Capture into resonance is only possible when the drift rate is below a critical value
that is approximately the square of
the inverse of the libration timescale; $|\dot b| \lesssim |\epsilon|^{4/3} |A|^{2/3}$,
corresponding to a critical planetary migration rate that is to order of magnitude
\begin{equation} 
\dot n_{crit} \sim \mu_p^{4/3} j^{5/3},
\end{equation}
where we have used the high $j$ (or $\alpha \to 1$) limit for the coefficient $\epsilon$.
Taking the square root of the momentum scale we can estimate the critical eccentricity
(below which capture has probability 1 in the adiabatic limit; \citealt{borderies84,malhotra90}) of 
\begin{equation}
e_{crit} \sim \left({\epsilon/A}\right)^{1/3} \sim \mu_p^{1/3} j^{-1/3}. \label{eqn:elim}
\end{equation}
This eccentricity value also corresponds to a mean value for the size of the eccentricity jump
that occurs
when the system crosses the resonance instead of capturing into it \citep{quillen06,mustill11}.
By differentiating the above two equations with respect to $j$, we can estimate the difference between 
critical drift rates for neighboring first order resonances,
\begin{equation}
{1 \over\dot n_{crit}} {d\dot n_{crit}\over dj} = {5 \over 3j}
\end{equation}
and critical eccentricity
\begin{equation}
{1 \over\dot e_{crit}} {d\dot e_{crit}\over dj} = -{1 \over 3j}.
\end{equation}
For $j =7$ the fractional difference in critical drift rates between the 7:6 and 6:5 resonances,
is of order 1/4 and is small. 
This is consistent with Figure 11 by \citet{mustill11} illustrating critical
drift rates for first and second order resonances.  

The fraction difference between the critical eccentricity
allowing capture into the 7:6 but not 6:5 is even more severe; it is only 5\%.
It would be difficult to maintain the system at a particular initial eccentricity value as
the two planets undergo secular oscillations as they migrate inwards, eccentricity damping
is likely during migration and as the system crosses resonances, planets can increase in eccentricity.
First order mean motion resonances also contain a corotation term that depends on the argument
\begin{equation}
\phi' = j \lambda - (j-1) \lambda_p + \varpi_p,
\end{equation}
and when the planet's eccentricity is similar to or above the critical eccentricity, this term can prevent capture
into resonance \citep{quillen06}.
For the Kepler 36 planets, with planet to stellar mass ratio listed in Table \ref{tab:tab1},
the critical  eccentricity is approximately $e_{crit} \sim 0.03$.   
 If the eccentricity damping rate is low, then we expect the planet 
eccentricities to increase if the system crosses first order resonances without capturing.
When the particle eccentricity is initially above the critical eccentricity value 
($\Gamma \gtrsim \Gamma_{res}$), the probability of capture into resonance
is not zero but is reduced.   Studies of stability have shown that higher eccentricity 
two body systems are often less stable than low eccentricity systems (e.g., \citealt{kley04,mustill11}). 
The 7:6 resonance is so close to the region of resonance overlap at low eccentricity
that even moderate planetary eccentricities put the system within the chaotic zone.
Given the number of processes affecting planet eccentricity, it would be extremely unlikely that a body could
remain near the critical eccentricity value for the 7:6 resonance as the system drifted inward, approaching the  resonance. 

We conclude that fine tuning is required to adjust either the migration rate or planet eccentricities 
so that capture (by smooth convergent migration) 
into the 7:6 mean motion resonance is possible but not other nearby mean motion resonances.

\subsection{Stochastic migration}

Stochastic forcing of a planet by disk turbulence or planetesimal scattering
can prevent resonance capture or kick a body out of resonance 
\citep{zhou02,murrayclay06,rein09,ketchum11,rein13,pardekooper13}.
Stochastic forcing 
has been described in terms of the variance of the angular momentum change
per orbit or equivalently in terms of a diffusion coefficient in angular momentum or semi-major axis.
 A planet wanders in semi-major axis a typical distance  
$$\delta a \approx \sqrt{D_a t }$$ after a time $t$,
where stochastic forcing is described in terms of a diffusion coefficient coefficient, $D_a$.
If the diffusion causes a random walk in semi-major axis, the above equation gives the standard deviation
of the distribution of distances travelled.
For $a$ in units of the planet's semi-major axis, $a_p$, and $t$ in units of $n_p^{-1}$ (the inverse of
the planet's mean motion), $D_a$  is in units of $a_p^2 n_p$. 
The diffusion coefficient, $D_a$, is approximately equivalent to the square of the $\alpha$
parameter adopted by \citet{rein13}, describing the ratio of the stochastic force perturbation per orbit in units
of the gravitational force from the central star.

A stochastically forced body escapes resonance
when the particle's semi-major axis varies by an amount of order the resonance width
 \citep{murrayclay06}. 
Equivalently, a variation in the coefficient $b$ (from our Hamiltonian; equation \ref{eqn:Ham}) 
greater than the resonance libration frequency would
 let the particle escape resonance;  
 \begin{equation}
 \delta b \sim j \delta n \gtrsim t_{lib}^{-1} ,
 \end{equation}
 for $\delta n$ the size of the variation in the mean motion of the planet. 
As long as the planet is at low eccentricity, this implies that the particle would escape resonance after 
a typical timescale, $t_{esc}$, with 
\begin{equation}
j  \sqrt{D_a t_{esc}} \sim |\epsilon|^{2/3} |A|^{1/3} \sim \mu_p^{2/3} j^{4/3}   
\end{equation}
giving a timescale for escaping the resonance
\begin{equation}
t_{esc} \sim D_a^{-1} \mu_p^{4/3} j^{2/3}, \label{eqn:tesc}
\end{equation}
where we have used equation \ref{eqn:tlib} for the libration timescale.

\subsection{Timescale to escape resonance via stochastic forcing when at the equilibrium eccentricity in resonance}


Once captured in resonance, the body's eccentricity increases with growth rate dependent on 
the drift rate.   The body's eccentricity increases until it reaches an equilibrium value $e_{eq}$, where
there is a balance between eccentricity growth
due to the resonance and eccentricity damping from dissipation forces \citep{gomes98,lee02}.
As the resonant width increases with eccentricity, the above relation (equation \ref{eqn:tesc}) 
for the escape timescale 
is a lower limit as it was estimated at low eccentricity.   To estimate the escape timescale after capture into
resonance we first estimate the equilibrium eccentricity, $e_{eq}$, and then we revise
our estimate for the escape timescale.

Once captured into resonance, the Poincar\'e momentum $\Gamma$ increases; 
equivalently the particle's eccentricity increases.
Hamilton's equation (using equation \ref{eqn:Ham}) gives 
$$\dot \phi = {\partial H \over \partial \Gamma }
 = 2 A \Gamma + b + {\epsilon\over 2} \Gamma^{-1/2} \cos\phi.$$
In resonance, the resonant argument librates about a fixed value so the average 
$\langle \phi \rangle $ (averaged over a libration or oscillation timescale) is constant
and $\langle \dot \phi \rangle = 0$.
As $\Gamma$ increases, the
first two terms on the right hand side of the above equation dominate over the third.   
Using these two approximations,
the time derivative of Hamilton's equation 
gives $2 A \dot \Gamma + \dot b =0$ and so an increase rate in the Poincar\'e momentum 
\begin{equation}
\dot \Gamma \approx {\dot b \over 2 A} . 
\end{equation}
Using $A\sim -j^{2}$ and a low eccentricity approximation for the Poincar\'e momentum,
the eccentricity growth rate in resonance is of order
\begin{equation}
e \dot e \sim {1 \over  j \tau_a}, \label{eqn:edot}
\end{equation}
where we have written the migration rate (setting $\dot b$) in terms of a migration timescale 
\begin{equation} \tau_a \equiv a/\dot a.  \label{eqn:taua}
\end{equation}
Equation \ref{eqn:edot}
implies that $e^2 \propto t$ in resonance and is consistent with previous work (\citealt{mustill11}; their equation 6).
Eccentricity damping on a timescale, 
\begin{equation} \tau_e = e/\dot e,  \label{eqn:taue}
\end{equation}
 can balance the eccentricity growth rate given by equation \ref{eqn:edot}.
This occurs at an equilibrium eccentricity that is estimated by balancing the two rates for $\dot e$; 
\begin{equation}
e_{eq} \approx \sqrt{\left|{\tau_e \over  j \tau_a }\right|} = {1 \over \sqrt{ j K}},\label{eqn:ebalance}
\end{equation}
where we have used a parameter
\begin{equation} K \equiv \tau_a/\tau_e \end{equation}
that describes the ratio of the eccentricity damping to
migration rate.
As is true in the case of dust particles captured into resonance by Poynting Robertson drag, 
for the high $j$ resonances, the equilibrium eccentricity is lower than for lower $j$ resonances \citep{liou97}.
For high rates of eccentricity damping (high $K$), the resonant system does not reach
as high eccentricities and so is more stable \citep{lee02,ketchum11}.  

When the equilibrium  eccentricity in resonance is higher than that set by 
dimensional analysis, or $\Gamma \gtrsim \Gamma_{res}$, 
the Hamiltonian can be approximated by\footnote{See the appendix on the Andoyer Hamiltonian by \citet{ferraz07}.} 
$$
H = Ap^2 + b' p + \epsilon \Gamma_0^{1/2} \cos \phi, 
$$
where $\Gamma_0 = \langle \Gamma \rangle$ is the average value of 
$\Gamma$, we have defined a new momentum, $p = \Gamma - \Gamma_0$, and we have incorporated the shift
in the distance to resonance with a 
new coefficient $b'$.
The resonant width is set by the mean eccentricity or $\Gamma_0 \approx e_{eq}^2/2$.
In this limit the inverse of the resonant libration frequency is 
\begin{equation}
t_{lib} = 1/\sqrt{|A\epsilon| \Gamma_0^{1/2}} \sim { j^{-3/2} \mu_p^{-1/2} e_{eq}^{-1/2}}.  \label{eqn:tlib_b}
\end{equation}
This gives a timescale to escape resonance (from $\delta b = j \sqrt{D_a t_{esc}} = t_{lib}^{-1}$) or
\begin{equation}
t_{esc} = {A\epsilon \Gamma_0^{1/2} \over j^2 D_a}  = {j \mu_p e_{eq} \over D_a}.
\end{equation}
Inserting equation \ref{eqn:ebalance} for the equilibrium eccentricity in resonance we estimate
an escape timescale
\begin{equation}
t_{esc} \sim {j^{1/2} \mu_p \over K^{1/2} D_a}. \label{eqn:tesc_balance}
\end{equation}
This timescale should be a better estimate than the timescale given in equation \ref{eqn:tesc}
as it takes into account the eccentricity of the body in resonance. 

Ignoring the weak dependence on $K$ and $j$ in equation \ref{eqn:tesc_balance}, the timescale to escape
resonance diffusively is approximately $t_{esc} \sim \mu_p D_a^{-1}$.  The stochasticity parameter
assumed by \citet{rein13} of $\alpha = 10^{-6}$, corresponding to $D_a \sim 10^{-12}$,
gives an escape timescale only of order $\sim 10^6$ orbital periods or $\sim 10^5$ years
taking into account the orbital period of the Kepler 36 planets. 
For the stochastic migration to account for the proximity of the Kepler 36 planets, 
the stochastic forcing would have had to dissipate on a  short timescale, otherwise
stochastic forcing would have removed the system from resonance.
\citet{kley04} pointed out a similar fine tuning problem.  A system that continues to migrate
after resonance capture can become unstable as the planetary eccentricities increase.
Here the system that continues to be stochastically forced after capture into resonance
can become unstable.    In the situation discussed by \citet{kley04}, the gaseous disk,
responsible for planetary migration must dissipate on a timescale short compared to the evolution timescale.
Here, the  source of the stochastic forcing must dissipate on a timescale short compared to the
evolution timescale of the system.

The  escape timescale estimated in equation \ref{eqn:tesc_balance} 
represents an estimate for the time that a stochastically forced
pair of planets in resonance would escape the resonance. 
This estimate neglects the possibility that dissipative forces (such as eccentricity damping)
could continually nudge the system into a stable region and the possibility that the stochastic
variations do not cause a true random walk in the semi-major axis.  
These factors might increase the lifetime of a resonant but stochastically forced system.
Systems in resonance may be unstable under gravitational forces alone and this might
decrease the lifetime of the resonant system.
We will discuss these issues further after we illustrate the behavior of stochastically forced systems numerically.

\subsection{Numerical Integrations}

We explore stochastic migration scenarios by integrating 
a few body system under the influence of gravity (a few planets and the central star) and including
a Stokes drag-like form for dissipation that induces both migration and eccentricity damping. 
The  drag gives a force per unit mass in the form adopted by \citet{beauge06}
\begin{equation}
{\bf F}_{drag} = - {{\bf v} \over 2 \tau_a} - {{\bf v - v}_c \over \tau_e},
\end{equation}
where $\bf v$ is the planet velocity and ${\bf v}_c$ is the velocity of a planet in a circular orbit
at the current radius (from the star) of the planet.
We use a 4th order adaptive step-size Hermite integrator (that described by \citealt{makino92}) with the addition of
the above drag force. 

We also introduce random velocity variations in the orbital plane (as did \citealt{ketchum11}). 
The distribution of velocity kicks is described by a Gaussian distribution with a standard 
deviation, $e_{kick}$, in units of 
the speed of a particle in a circular orbit at the current position of the planet.  
The velocity kicks are given twice per orbit.  
The diffusion coefficient discussed above; $D_a \sim e_{kick}^2$.
Migration rates, eccentricity damping timescales and stochastic forcing parameters $\tau_a, \tau_e, e_{kick}$
are set individually for each massive body in the integration except the central star.

During the integration, we continually check for collisions between bodies (planets, embryos or the star). 
For most of our integrations
we use the measured masses and radii of the planets Kepler 36b and c
and its host star.   
A collision is identified when the two bodies
have distance between
their center of masses that is equal to or within the sum of their radii.
Following \citet{asphaug10}, at the moment of impact,
we define the angle of impact as the angle between between the relative velocity vector and
the vector between the two center of masses;
\begin{equation}
\theta_{im} = {\rm acos}\left( {- {\bf v}_{impact} \cdot \Delta {\bf r} \over |{\bf v}_{impact}|| \Delta {\bf r}|} \right).
\label{eqn:theta_im}
\end{equation}
Here $\theta_{im} = 0$ for a direct normal collision and $\theta_{im}=90^\circ$ for a grazing collision where
the surfaces barely touch.
A grazing impact is defined as one with angle such that the center of mass of the smaller body would graze  the surface of the larger body,
\begin{equation}
\theta_g \equiv \sin^{-1} \left({ R \over R+r}\right), \label{eqn:theta_g}
\end{equation} 
where $R$ is the radius of the larger body and $r$ is the radius of the smaller body.
If $\theta_{im} > \theta_g$ then the surface of the larger body can be removed during the collision 
\citep{asphaug10}.

We work in units of the innermost planet's initial semi-major axis,  
and the mass of the central star.  Time is such that the innermost planet's initial orbital period has a value
of $2 \pi$.  We primarily plot figures in units of the innermost planet's initial orbital period.

\subsection{Illustrations of stochastic migration for two planets}

\begin{figure}
\includegraphics[width=3.5in,trim=0.1in 0.1in 0.1in 0.1in]{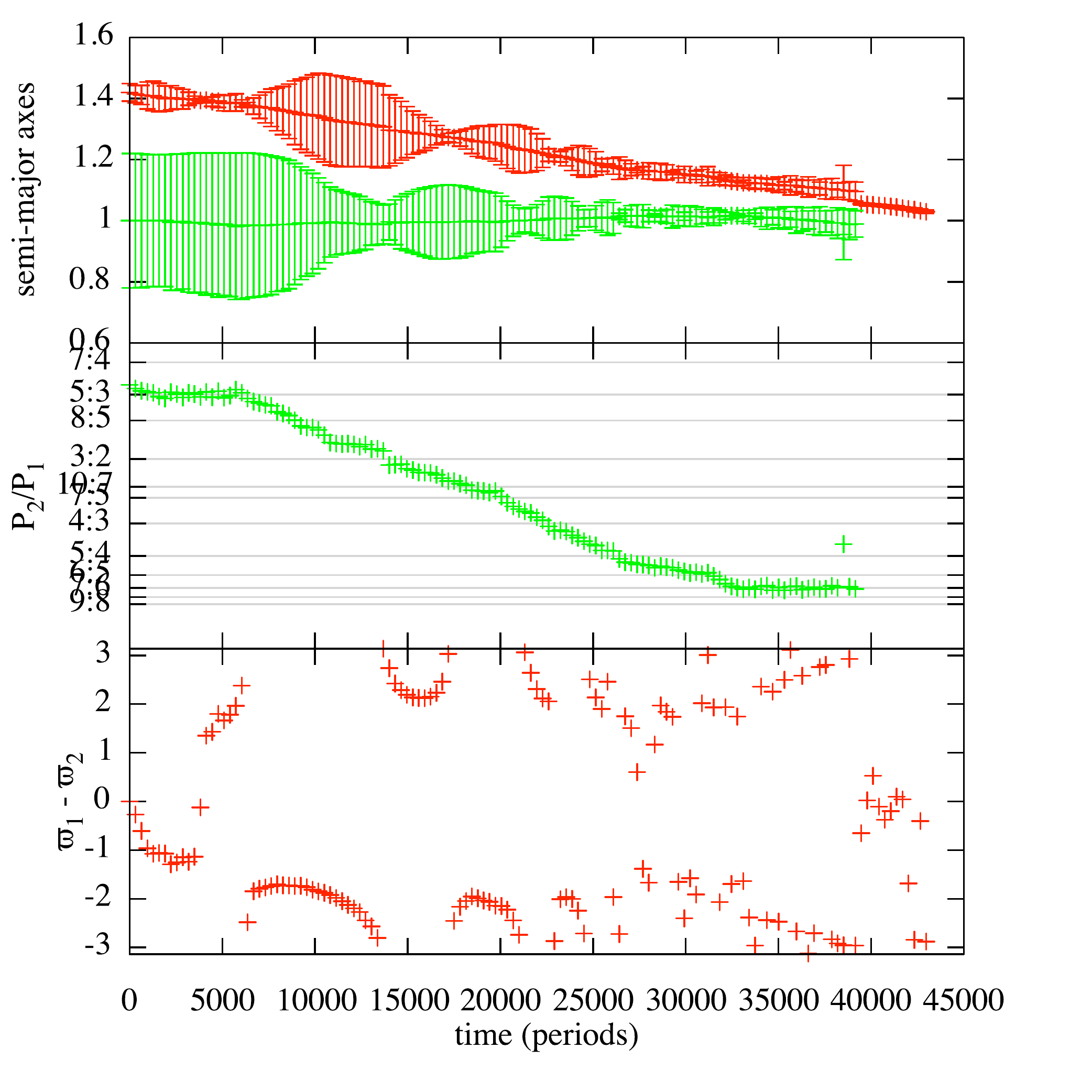}
\caption{
Integrations of two planets, where 
 the outer planet migrates stochastically with migration timescale
$\tau_a = 10^6$, eccentricity damping factor $K=20$, and stochasticity parameter $e_{kick} =  10^{-5}$.
The planets have masses and radii of those measured for Kepler 36b and c.
The inner planet was initially started with a moderate eccentricity of 0.2.
The top panel shows semi-major axes, with error bars set according to the planet's eccentricity
so that they illustrate pericenters and apocenters of the orbit as a function of time.
The middle panel shows the ratio of orbital periods and major mean motion resonances are shown
with horizontal lines.
The bottom panel shows the difference between the planet's longitudes of pericenter.
The eccentricity variations are due to secular oscillations.
The capture probability is reduced both
by the eccentricity and the stochasticity.   If the eccentricity, migration rate and stochasticity of the planets
is sufficiently high then low $j$ resonances can be bypassed.  In this simulation the planets were
captured into the  7:6 resonance, however the system did not remain there for long. \label{fig:x}
}
\end{figure}

\begin{figure}
\includegraphics[width=3.5in,trim=0.1in 0.1in 0.1in 0.1in]{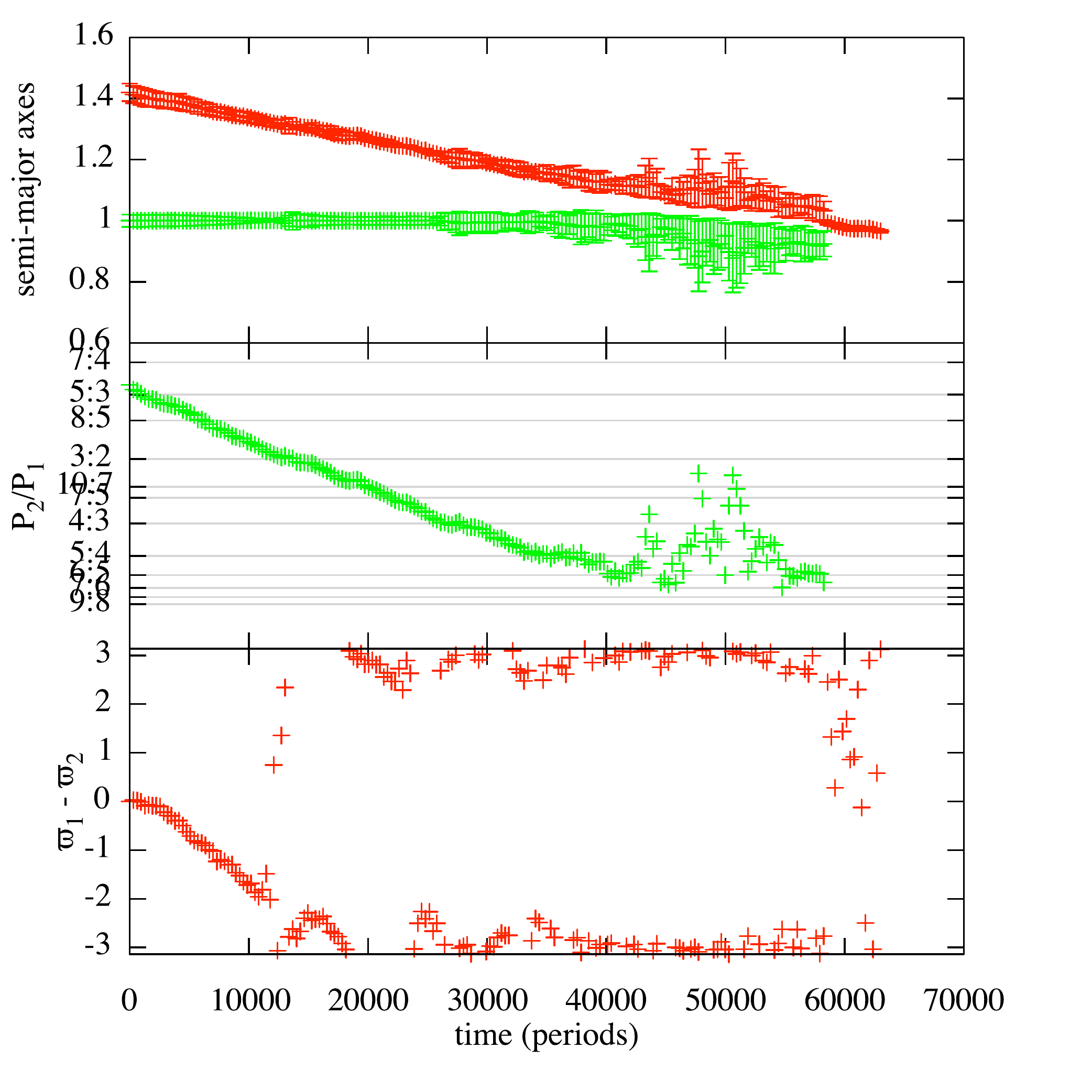}
\caption{
Similar to Figure \ref{fig:x} except
 both planets had initially low eccentricities of 0.02 and the stochasticity parameter
 $e_{kick} = 1.5 \times 10^{-5}$.
After 40,000 orbital periods, the two planets experienced an era of strong close 
encounters that ended with a collision. \label{fig:x2}
}
\end{figure}

Example stochastic migration integrations for two planets 
are shown in Figures \ref{fig:x},  and \ref{fig:x2}.  
For both integrations, the two planet masses and radii are equivalent to those measured for Kep b and c.
The outer planet's initial semi-major axis is 1.42 times that of the inner planet.
The forced migration timescale $\tau_a = -10^6$ for the outer planet and is infinity for
the inner planet.  For the outer planet $K=20$, setting the eccentricity damping timescale.  
Estimates of the ratio of the eccentricity damping ratio, $K \equiv \tau_e/\tau_e$, from hydrodynamic simulations
range from $K \sim 1$ \citep{kley04}, 10 \citep{cresswell08} to 100 \citep{bitsch10}.   We have chosen
values for $K$  in the middle of this range.
The stochasticity parameter $e_{kick} = 10^{-5}$ and $1.5 \times 10^{-5}$ for Figures \ref{fig:x} and \ref{fig:x2},
respectively.  Initial orbital angles were randomly chosen.   The initial inclinations were extremely small but
nonzero; the outermost planet has an initial inclination of 0.1$^\circ$.

In Figure \ref{fig:x} and \ref{fig:x2},
the top panel shows semi-major axes, with error bars set according to the planet's eccentricity
so that they illustrate pericenters and apocenters of the orbit as a function of time.
Both planets are shown together in this panel.
The middle panel shows the ratio of orbital periods and major mean motion resonances are shown
with grey horizontal lines.
The bottom panel shows the difference between the planetÕs longitudes of pericenter and illustrates when the two bodies are in resonance.\footnote{The perturbation terms in the Hamiltonain with resonant arguments 
$\cos \phi$ and $\cos \phi'$ have 
opposite sign.  When the different between the planets' longitudes of pericenter is $\varpi_b - \varpi_c \approx \pi$,
the two terms constructively add and the resonance is effectively stronger.} 
In both integrations, the time between reaching the vicinity of the (or capture into) 7:6 resonance and 
a planet/planet collision  was short compared to the migration timescale.
The two integrations shown were chosen from approximately 30 integrations with different initial orbital angles
planet eccentricities,  $\tau_a, K$ and $e_{kick}$ values.   The two integrations shown were chosen because
the two planets remained the longest time near or in the 7:6 resonance.
When we adjust the stochasticity parameter so that lower $j$ mean motion 
resonances are bypassed, we find that the system is unstable and collisions between the two planets are likely, on a short timescale, independent of whether the planets are begun 
at low or high initial eccentricity.  The short lifetime of stochastically forced planets in the 7:6 resonance
we see in the integrations
is consistent with the short timescale we have estimated analytically for the resonant lifetime. 

Both our order of magnitude calculations and our numerical integrations suggest that capture into the 7:6 resonance is  possible for stochastically forced migrating planets, but that the planets are unlikely to remain there for long.   
This is not confirmed by the hydrodynamic study (and accompanying thorough N-body study) by 
\citet{pardekooper13} who find that the planets are likely to be placed into and remain in stable resonant
configurations despite the stochastic forcing.   It may be possible to reconcile our order of magnitude
estimates with their results by  improving upon our simplistic random walk prescription of
the diffusive process.

\section{Stochastic forcing by planetary embryos}

We consider the possibility that stochastic forcing is associated with encounters with
planetary embryos, bodies of order Mars or Earth mass (as previously studied by \citealt{liu11}), 
rather than turbulence from a gaseous disk.
A disk edge can trap embryos forming a `planet trap' \citep{masset06} where planetary embryos
can collide with each other \citep{morbi08} or with planets in the vicinity of the disk edge.
A feature of this scenario is that collisions between embryos and planets are potentially capable
of stripping the outer layer of a planet and so account for the high density of Kepler 36b. 

We consider a group of  identical mass planetary embryos, 
with planet to stellar mass ratio, $\mu_{embryo}$
initially exterior to two planets.  
We integrate the system allowing the planetary embryos and the outermost planet to migrate inwards.
Stochastic forcing, (due to turbulence in a gas disk) is set to zero for all bodies.  
The neglect of stochastic forcing due to a turbulent disk 
reduces the number of free parameters, and is consistent with a low value, 
$\alpha \lesssim 10^{-6}$,
of the poorly constrained (see \citealt{rein12,nelson04}) turbulent forcing parameter. 

As did \citet{morbi08}, we allow the embryos to migrate faster
than the planets.   This was motivated by comparing a type I migration rate (appropriate for embryos embedded
in a disk) to a type II migration rate (appropriate for gap opening planets), within the context of torques transferred
between embryos or planets to a gaseous disk via driving of density waves \citep{ward97,tanaka02}.
If and when the two planets exchange locations so
the inner planet becomes the outer one, then the migration and eccentricity damping rates are also
 swapped.   This would be expected if the outer planet opened a gap in the gaseous disk,
 and the gas did not significantly penetrate within its orbit.  
This procedure allows convergent migration at all times for the two planets.
In the event of a collision, momentum and mass were conserved and a new larger body created 
at the location of the center of mass of the collision with a density equivalent to the 
mean density of both progenitors.

We ran sets each with 10 numerical integrations that we denote the X-series, Y-series, A-series and B-series.
The X-series set has planet masses and radii equivalent to those of the Kepler 36 b and c.
The Y, A and B-series sets have two equal planet masses with mass and radius equivalent to Kepler 36c.  
Parameters for these integrations 
are listed and described in Table \ref{tab:tab2}. 
The X and Y-series integrations have $\mu_{embryo} = 7 \times 10^{-7}$ (equivalent to 2.5 the mass of Mars). 
The mass ratio $\mu_c/\mu_{embryo} = 30$ is sufficiently large that collisions can be disruptive \citep{benz07,asphaug10,stewart12}.  The A and B-series simulations have lower embryo masses, with 
$\mu_{embryo} = 4 \times 10^{-7}$
and $\mu_{embryo} = 1 \times 10^{-7}$, respectively.   Embryo to planet mass ratios (52 and 210) for these
simulations would require impact velocities significant above the circular velocity to be high disruptive during
a planet/embryo collision.
X-series and B-series integrations were run with 7 embryos only. 
Y-series and A-series integrations were run with 4, 7 or 10 embryos (and 10 integrations in each case).

\begin{table}
\vbox to 140mm{\vfil
\caption{\large Integrations with Two planets and  Embryos 
     \label{tab:tab2}}
\begin{tabular}{@{}lcccc}
\hline
Integration     series                 &  X-series                          &  Y-series &  A-series   & B-series\\
$\mu_b/10^{-5}$                                 &   $1.15 $    &  $   2.09  $     & $   2.09 $ &$   2.09  $  \\
$\mu_c/10^{-5}$                                 &   $ 2.09 $   &   $   2.09 $   & $   2.09   $  & $   2.09  $  \\  
$R_b$                                     &  0.00054                          & 0.001  & 0.001 &0.001 \\
$\mu_{embryo}/10^{-7} $                     &   $7 $   & $7 $  & $4 $ & $1 $ \\
$R_{embryo}$                           &  0.0003  & 0.0003 & 0.00027 & 0.00017\\
\hline
$R_c$                         &  \multicolumn{2}{c}{ 0.001 }\\  
$\tau_a$   Outer planet         &\multicolumn{2}{c}{$2 \times 10^{6}   $}          \\
$\tau_a$  Embryos                &\multicolumn{2}{c}{ $10^6$}                          \\
$K$   Outer planet                 & \multicolumn{2}{c}{20 }                                  \\     
$K$  Embryos                        & \multicolumn{2}{c}{10}                                 \\
   \hline
\end{tabular}
{\\ Here $\mu_b,\mu_c,\mu_{embryo}$ refer to the body to stellar mass ratio for the inner planet, outer planet and embryo masses, respectively.  The parameters $R_b, R_c, R_{embryo}$ are the body radii in units of
the initial semi-major axis of the innermost planet.  
The embryo radius was chosen so that embryos have the same mean density as Kepler 36c. 
The innermost planet has no forced migration.   There is no stochastic forcing in these integrations.
The initial semi-major axis ratio of the two planets is $a_c/a_b = 1.4$.  The embryos were initially separated so
that the ratio of the semi-axis to that of the next nearest body is $a_{i+1}/a_i = 1.1$.  Embryos were begun on circular
orbits exterior to the planets.  Initial eccentricities of all bodies were set to zero.  
The parameter $\tau_a$ describes a timescale
for migration and is given in units of the inverse of the inner planet's initial mean motion.
The parameter $K$ is the ratio of the eccentricity damping rate to the migration rate.
Initial orbital angles were chosen randomly.  Initial inclinations were chosen randomly and within 0.1 degrees
of the mid plane. Integrations were run for a time of between 200000 and 400000 orbital periods 
(of the inner planet initially) allowing all inward migrating embryos to interact with the planets.
Five of these integrations are shown in Figures \ref{fig:exchange}, \ref{fig:chain}, \ref{fig:76} and \ref{fig:number}.
}
\vfil}
\end{table}

From the 10 X-series of integrations with 7 embryos 
the end states were 4 integrations with collisions between the two planets, 
though at the end of one of these an embryo was trapped in the 5:4 resonance with the remaining planet,
one integration each with planets in the 4:3, 5:4, 7:5 and 6:5 resonances, one simulation with the planets near the 7:5 resonance but at moderate eccentricity so likely to experience a collision later on, and a resonant chain with
the two planets in the 4:3 resonance and an embryo in a 3:2 resonance with the outer planet.
In two integrations of these integrations, the two planets swapped locations.
From the 10 Y-series of integrations with 7 embryos
the end states were 3 integrations with two planets in the 4:3 resonance,
one integration with planets in the 5:4, three in the 6:5, one in the 7:5 and one in the 7:6 resonance.
The final result of one integration was a resonant chain with the inner two planets in a 7:5 resonance
and the remaining embryo in a 4:3 with the outer planet.  In the Y-series with 7 embryos
there were no planet/planet collisions
nor did the planets swap locations.
From the 10 A-series of integrations with 7 embryos, end states were two planet/planet collisions,
two integrations with planets in each of the 7:5, 5:4 resonances and 4 integrations in the 4:3 resonance.
In one of the simulations after a planet/planet collision two embryos were left in the 5:4 resonance
with the planet (and they were corotating).
From the 10 B-series of integrations with 7 light embryos, all simulations ended with the two planets
in the 3:2 resonance.

\begin{figure}
\includegraphics[width=3.5in,trim=0.1in 0.1in 0.1in 0.1in]{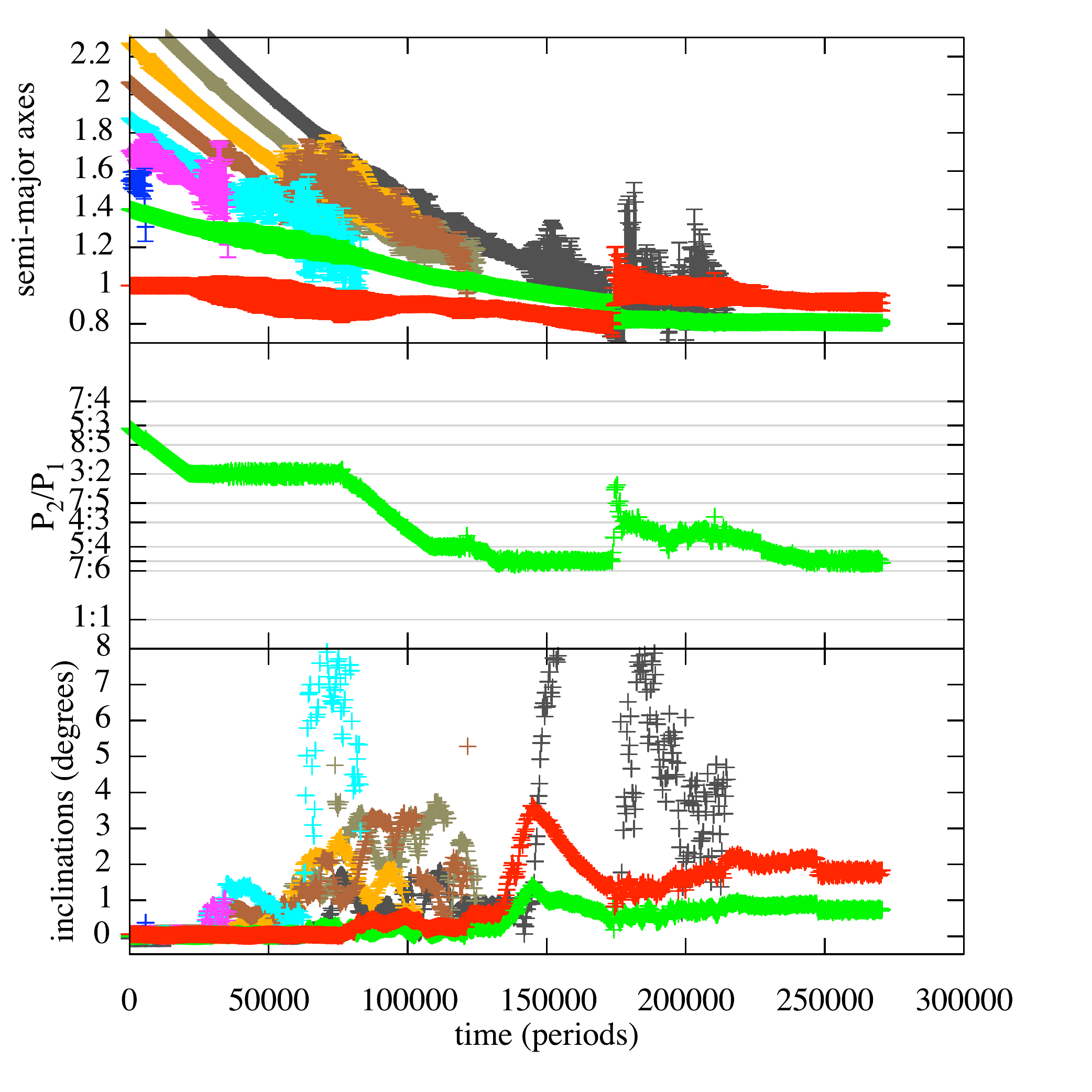}
\caption{
Stochastic forcing from planetary embryos can knock two planets out of resonance and in some cases
allows them to swap locations.  The integration displayed is from the X-series with parameters shown
in Table \ref{tab:tab2}.
The top panel shows semi-major axes, with error bars set according to the body's eccentricity
so that they illustrate pericenters and apocenters of the orbit as a function of time.
Here the red and green points are the two planets and the other color points correspond to planetary embryos.
The middle panel shows the ratio of orbital periods for the two planets (outer divided by inner)
 and major mean motion resonances are shown
with horizontal lines.
The bottom panel shows inclinations of the bodies, with the same color points as the top panel.
Collisions with the outermost planet (green points) took place before $t=1.25 \times 10^{5}$  periods
except  for the last collision that took place at $t=2.1 \times 10^5$ and with the same body after it
had exchanged places with the other planet (red points).
This integration illustrates that the positions of the two planets in some cases exchange or swap locations
and leave the two planets in adjacent orbits.
The planet originally closer to the star experienced no
collisions with embryos.   At the end of the simulation, the planet that experienced collisions is the innermost planet.  If this planet was stripped during a collision, a dense core could have been left behind.   This scenario could
account for the large density contrast between the Kepler 36 planets.
\label{fig:exchange}
}
\end{figure}

\begin{figure}
\includegraphics[width=3.5in,trim=0.1in 0.1in 0.1in 0.1in]{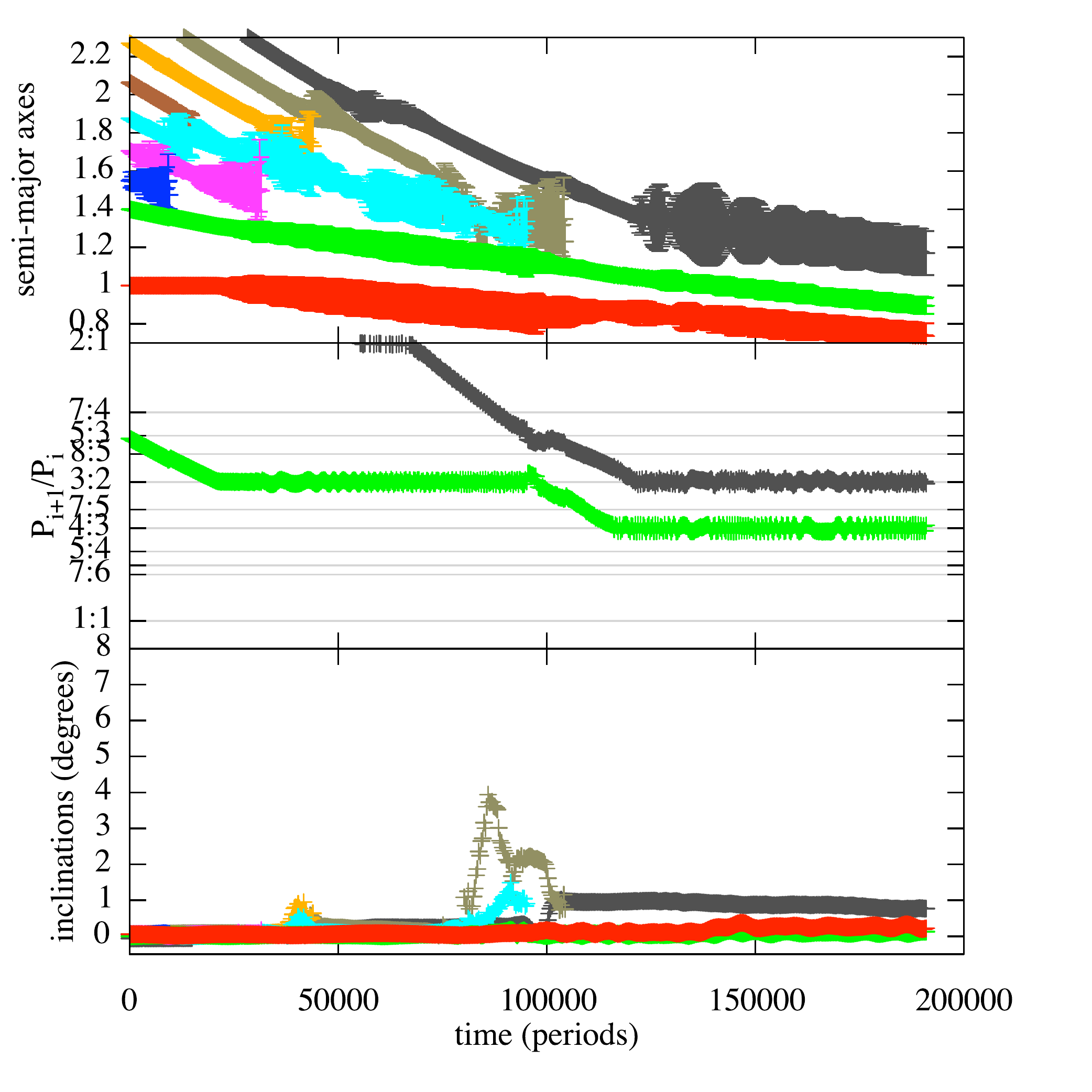}
\caption{
Similar to  Figure \ref{fig:exchange} except illustrating a different simulation from the
X-series with 7 embryos.  This integration illustrates that a resonant chain can be the end state. 
The middle panel shows that at the end of the simulation the two planets have the period ratio of the 
 7:5 resonance and the remaining embryo is in the 3:2 resonance
with the outer planet.
\label{fig:chain}
}
\end{figure}

\begin{figure}
\includegraphics[width=3.5in,trim=0.1in 0.1in 0.1in 0.1in]{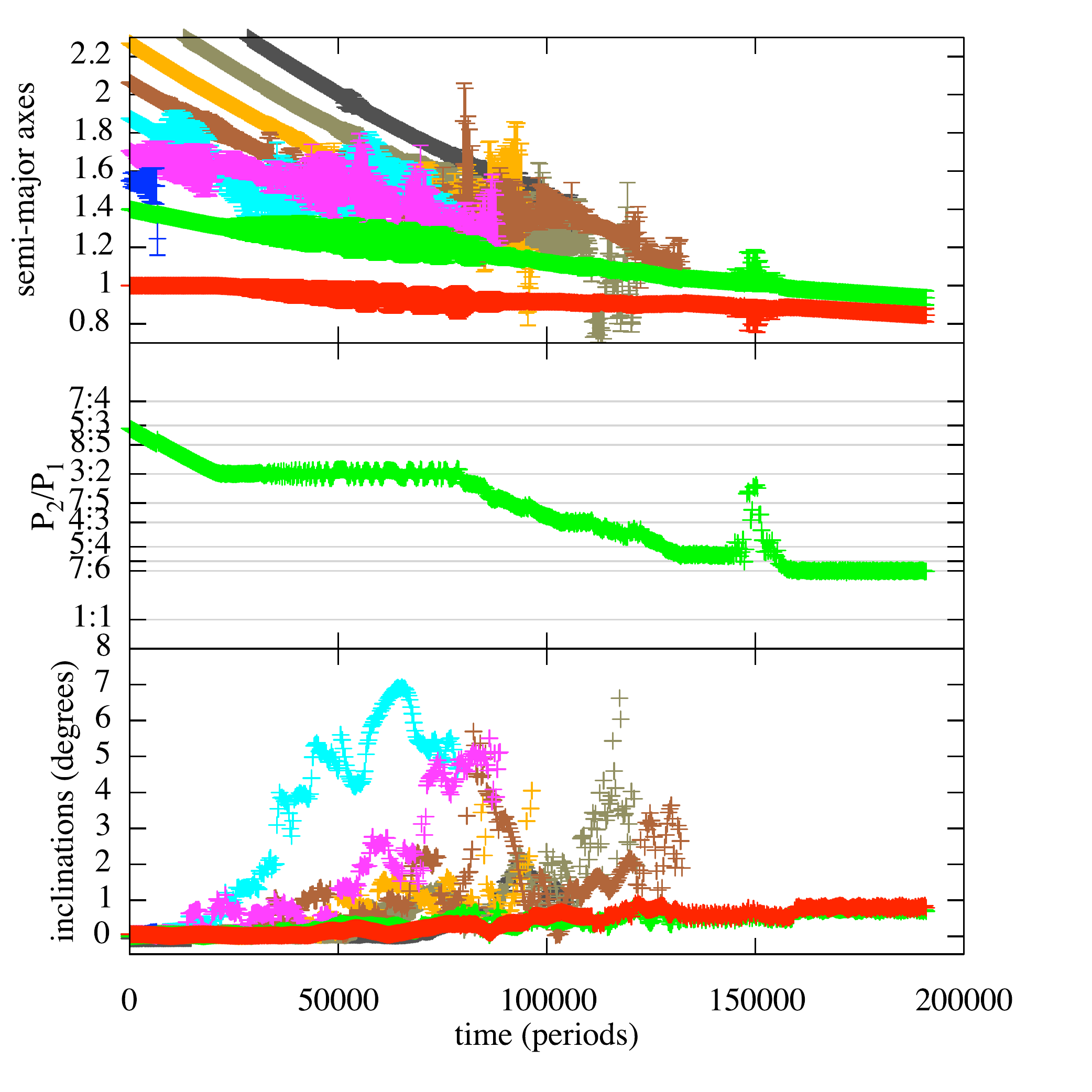}
\caption{
Similar to  Figure \ref{fig:exchange} except showing an integration from the Y-series with 7 embryos.
This integration illustrates that the final result can be two planets in the 7:6 resonance. 
The innermost planet experienced two collisions with two different embryos (grey and brown points) at
  $t \approx 1.2$ and $1.3 \times 10^5$ periods.
If the planet was stripped during a collision, a dense core could have been left behind, accounting for the high density of Kepler 36b.
\label{fig:76}
}
\end{figure}

Figures \ref{fig:exchange} and \ref{fig:chain} show integrations from the X-series and Figure \ref{fig:76} shows
an integration from the Y-series, both with 7 embryos.
The top panel shows semi-major axes, with error bars set according to the body's eccentricity
so that they illustrate pericenters and apocenters of the orbit as a function of time.
Here the red and green points are the two planets and the other color points correspond to the planetary embryos.
The middle panel shows the ratio of orbital periods for the two planets (outer divided by inner)
 and major mean motion resonances are shown
with horizontal lines.
The bottom panel shows inclinations of the bodies, with the same color points as the top panel.
Inclinations were measured with respect to the initial orbital mid plane.
These integrations illustrate that encounters with planetary embryos can
knock the convergently migrating planets out of resonances and so allow them to get closer together.
Gravitational encounters with embryos can serve as a type of stochastic forcing that pushes the system
past resonances such as the 4:3 and 5:4 resonances, but sometimes lets the planets into the 7:6 resonance. 

What mass embryo is sufficiently massive to push a planet out of resonance? 
Comparing the end states between the Y, A and B series, with decreasing embryo mass,
we found that the lowest mass embryos in the B-series, were not sufficiently massive
to knock two planets out of the 3:2 resonance.
We  estimate the size of a variation in semi-major
axis or equivalently energy that would remove a system from resonance.   
As we did previously, we estimate the width of the resonance  
 as $b \sim j \delta n \lesssim  t_{lib}^{-1}$ for a variation $\delta n$ from the center of the resonance.
 A variation in mean motion can be related to a change in semi-major axis and so energy.
Using equation \ref{eqn:tlib_b} 
 for the libration frequency (describing the resonant width
at an eccentricity $e_{eq}$), we estimate a variation in orbital energy of order
\begin{equation}
{\delta E \over E} \gtrsim j^{1/2} \mu_p^{1/2} e_{eq}^{1/2},
\end{equation}
would remove a planet from the $j:j-1$ resonance with another planet.
Using the impulse approximation we estimate that an embryo 
undergoing a gravitational encounter with impact parameter $b_{en}$ (in units of the planet's semi-major axis) 
with a planet would cause a change in orbital energy of order
\begin{equation}
{\delta E \over E} \sim { \mu_{embryo}   \over v_{en} b_{en}} 
\end{equation}
where the encounter velocity $v_{en}$ is units of the planet's circular velocity.
Equating these two relations we estimate that an encounter with $v_{en} b_{en}$ and embryo mass ratio
$\mu_{embryo}$
\begin{equation}
\mu_{embryo} \approx j^{1/2} \mu_p^{1/2} e_{eq}^{1/2} v_{en} b_{en}
\end{equation}
can remove the planet from resonance.
The minimum impact parameter is given by the radius of the planet and is of order 1/1000 in units
of the semi-major axis (see Table \ref{tab:tab1}).
Taking $v_{en} \sim 0.2$ the circular velocity (measured from the distribution of collisions between 
planets and embryos
as we will discuss below),  $b_{en} = 1/1000$,  and the mass ratio of Kep 36c, we estimate
that an embryo with mass ratio of order $\mu \sim 10^{-6}$ is required to knock the Kepler 36 system out
of a low $j$ first order mean motion resonance.
Convergent migration of the two planets allows them to be captured in resonances such as the 3:2 resonance
(see Figure \ref{fig:exchange}).   In most but not all cases, the planets escaped resonance because
of a collision with an embryo.    In some cases, close encounters were sufficient to kick the system out of resonance.
The embryo masses in our integrations that allowed the two planets to escape
resonances ($\mu_{embryo} = 4\times 10^{-7}$ and larger but not $\mu_{embryo} = 10^{-7}$) 
is consistent with this mass estimate.

We find that 
there is a diversity of possible end states exhibited by the integrations.
These include resonant configurations with two planets alone, including 
high $j$ resonances such as the 7:6 (see Figure \ref{fig:76}).  
This end state is in a fairly stable region,  though longer 
integrations that include a depletion timescale to remove the dissipative forces are needed to determine if
this end state would be long lasting and leave the two planets in a configuration consistent with
that observed for the Kepler 36 system.  The Kepler 36 system lies just outside the 7:6 resonance, so subsequent evolution would have to account not only for the stability of the system but removal from the resonance.
A couple of our integrations ended with a last embryo in a resonant chain configuration where each pair
of bodies is in resonance (see Figure \ref{fig:chain}).  
The Kepler planetary systems include resonant chains (e.g.,  the KOI 730 system \citealt{lissauer11}),
so the diversity of our integrations is not inconsistent with the diversity of Kepler planetary systems.
Two of the integrations show the two planets exchanging locations (see Figure \ref{fig:exchange}).  
This is reminiscent of the exchange of Neptune and Uranus in some `Nice' model simulations 
of the early Solar system evolution \citep{morbi10}.   Among our 20 integrations, we find that
 the outermost planet experiences approximately twice as many collisions 
with embryos as the innermost planet.   These collisions could have stripped the envelope of this planet, leaving it 
at higher mean density than the other one.  Because the two planets swap locations, the end result might be a high density inner planet  and a lower density outer planet, as is found in the Kepler 36 system.
The outer planet could continue to accrete gas, increasing its mass and lowering its mean density after the swap 
took place.  At the end of the simulations shown in Figure \ref{fig:number} an embryo is left interior to the two
planets.  This suggests a mechanism for moving a small body closer to a star, 
relevant for interpretation of small hot bodies such as the planet Kepler 37b \citep{barclay13}, though here
the three bodies at the end of the simulation are not coplanar and the three Kepler 37 planets
are all seen in transit and so are coplanar.  The integrations that end with two planets in the second order
7:5 resonance are potentially relevant for interpretation of systems near this resonance such as the two Super-Earths orbiting HD41248 \citep{jenkins13}.

These integrations illustrate that the `planet trap' setting, with numerous embryos
embedded in the gas disk exterior to two planets at the edge of a clearing in the disk,
is promising as it could
account for both the proximity and high density contrast of the Kepler 36 planets.


Transit timing observations can place constraints on orbital inclinations (as discussed by \citealt{carter12}).
When collisions between planets and embryos occur, planet inclinations can be excited.
For example, at the end of the integration shown in Figure \ref{fig:exchange}, the two planets differ by a degree
in their orbital inclination.    Future observations of the Kepler 36 system may determine if there is a difference in 
inclination between the two planets.  The resonant chain shown in Figure \ref{fig:chain} contains an exterior embryo
at a different inclination than the two planets.    It is possible that such an object could exist 
in the Kepler 36  system but is not detected in transit.  In the simulations shown in Figure \ref{fig:number}
a embryo was left interior to the two planets and at a differing inclination.  This object too would
not be seen in transit.
 At the end of the integration in 
Figure \ref{fig:76}, the two planets are at a different inclination than they were originally 
but we found that the two planets are are approximately coplanar 
(their mutual inclination is less than $0.2^\circ$), so
not all integrations ended with mutually inclined planets.
In this simulation no embryos were ejected so the variation in angular momentum 
is due to grazing encounters.   Our integrations do not conserve angular momentum during
collisions as we do not record or adjust planet rotation rates.

\subsection{Properties of collisions}

\begin{figure}
\includegraphics[width=3.5in,trim=0.1in 0.1in 0.1in 0.1in]{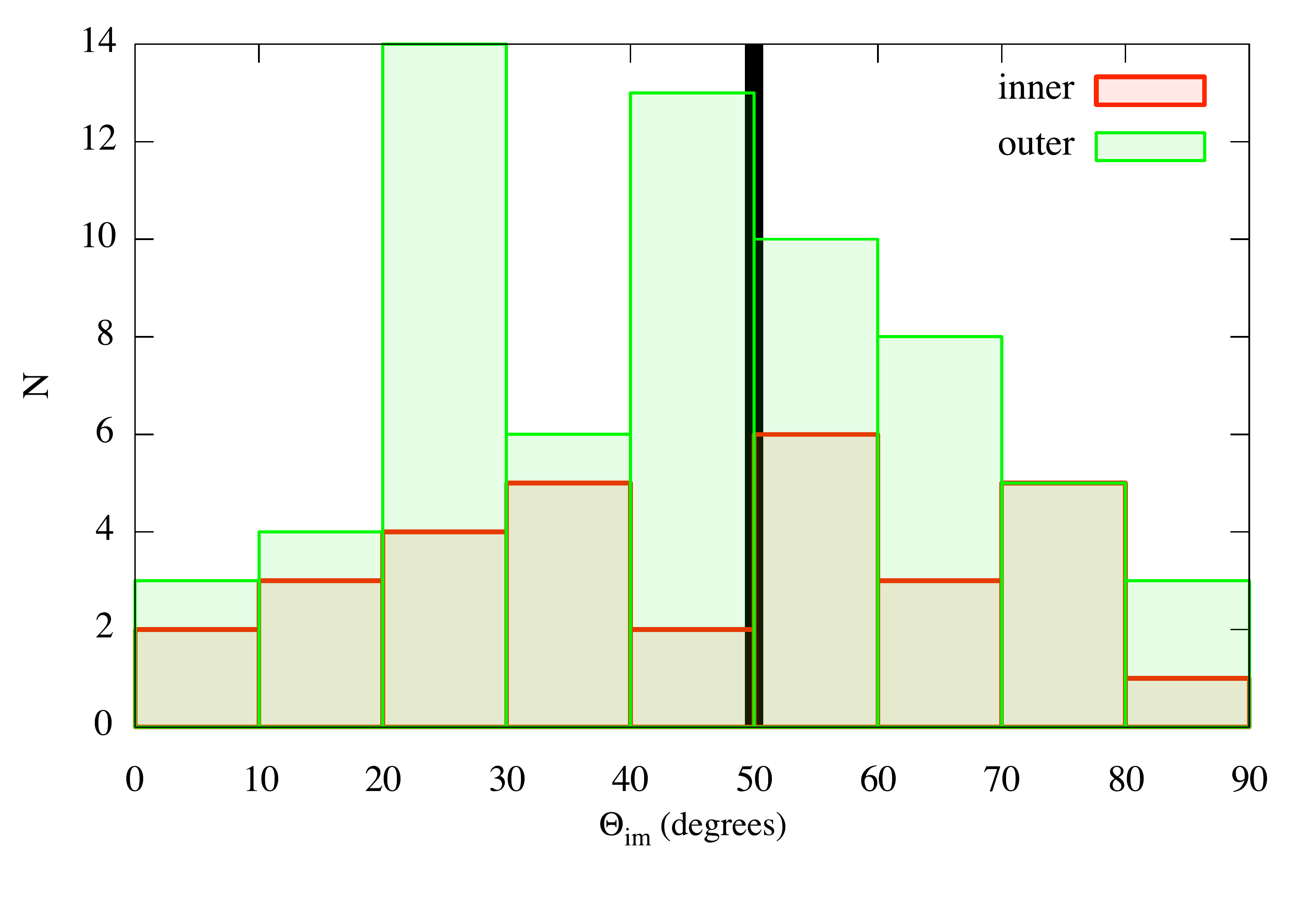}
\includegraphics[width=3.5in,trim=0.1in 0.1in 0.1in 0.1in]{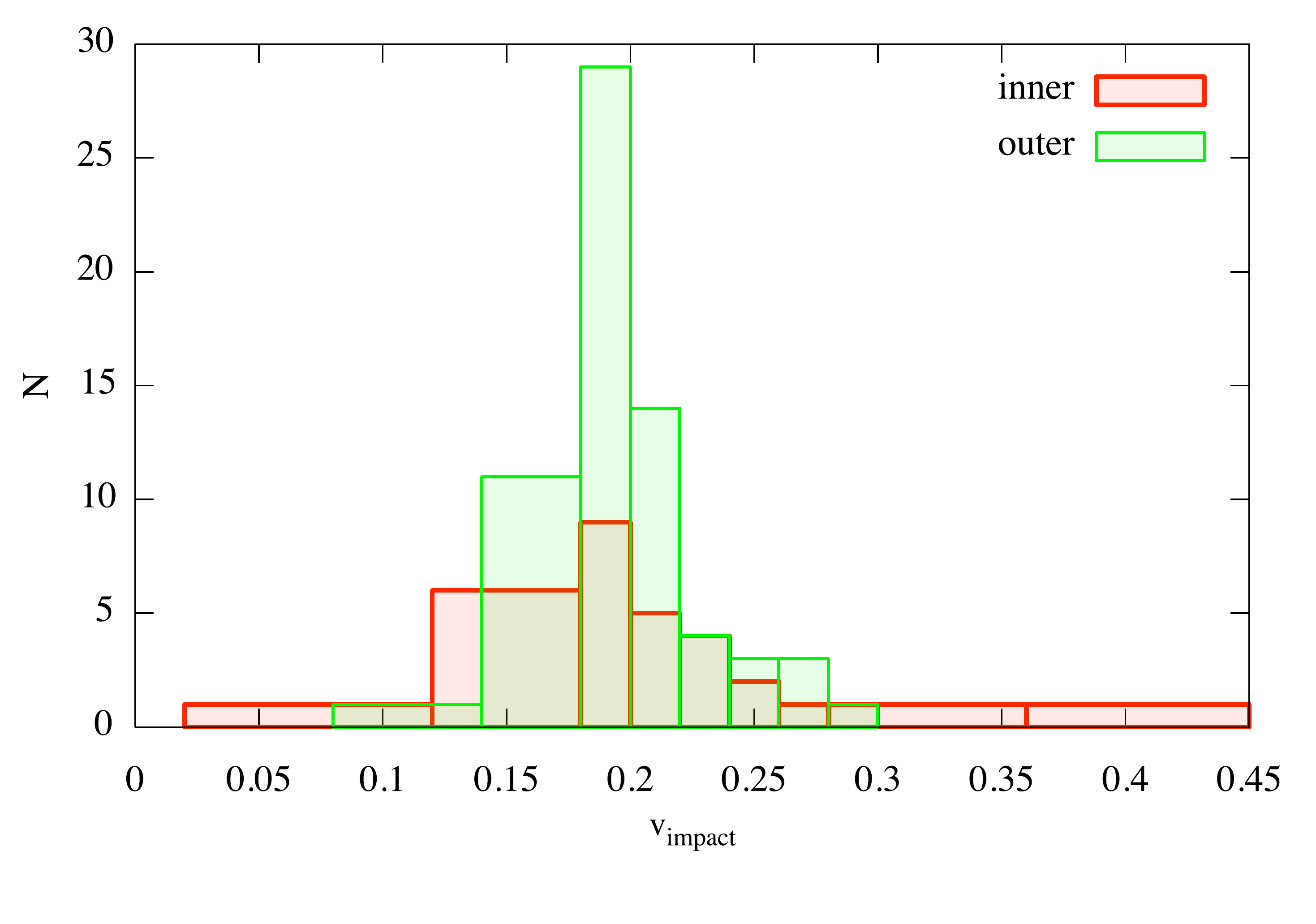}
\caption{
a) Distribution of collision impact angles measured from embryo/planet collisions
in the X-series and Y-series integrations with 7 embryos.  
Red  and green histograms show collisions with the innermost and 
outermost planets, respectively.
Here an impact angle of $0^\circ$ is a normal impact and an impact angle of 90$^\circ$ is grazing.
 Impact angles greater than $50^\circ$ are sufficiently
grazing that they can disrupt the envelope of a planet such as Kepler 36 c.
b) The distribution of impact velocities in units of the initial innermost planet's circular velocity. 
Here a value of 0.2 is corresponds to the escape velocity of the Kepler 36 planets.
 The velocity distribution is wider for the inner planet.  This is 
 expected as embryos are likely to be at higher eccentricity when they cross the orbit of the inner planet.
There is a wide distribution of impact angles and velocities implying that accretionary, disruptive,
and envelope stripping collisions are possible.
\label{fig:col}
}
\end{figure}

For all collisions in the X and Y-series of integrations with 7 embryos
we recorded the impact angles and relative velocities at impact of the two bodies involved in the collision.
Histograms for the collision properties are shown in Figure \ref{fig:col} for planet/embryo impacts.
The distributions are computed separately for each planet.
The velocities are given in units of a particle in a circular orbit with the 
innermost planet's initial semi-major axis.  The impact angle was computed with equation 
\ref{eqn:theta_im}.  
We compute the critical grazing angle, 
(equation \ref{eqn:theta_g}) using the radii
from Table \ref{tab:tab2}, finding $\theta_g = 50^\circ$ for impacts
between a planet with the radius of Kep 36c and a planetary
embryos with the same density but 1/28th the mass, finding $\theta_g = 50^\circ$. 
Collisions with $\theta_{im} \gtrsim \theta_g$ are capable of stripping the outer envelope of a planet
\citep{asphaug10}.  The critical impact angle of $\theta_g = 50^\circ$ is shown as a wide, black,
 vertical line in Figure \ref{fig:col}a.
The impact angle distribution with both planets is wide, implying that both normal and grazing impacts
occur with both planets.  

As shown in Figure \ref{fig:col}b, the impact velocity distribution is wider for the inner than outer planet.
This is expected
as embryos are likely to have higher eccentricities when they cross the orbit of the inner planet than the outer planet.
We have shown the impact velocity distributions in units of the circular velocity.   However  the escape velocity from the planets is about 5 times smaller than the circular velocity (see Table \ref{tab:tab1}).
The mean of the impact velocity distributions are at a value of 0.2 times the circular velocity and so are approximately
equal to the escape velocity of the planets.

Simulations of terrestrial body collisions have shown that the outcome of collisions in the gravity-dominated regime
can be described statistically with scaling laws that primarily dependent on impact velocity in units of the escape
velocity, impact angle and projectile to target mass ratio \citep{marcus10,leinhardt12}.
Planet constituents affect the scaling laws through a material parameter defining the catastrophic disruption criteria between equal-mass bodies in units of the specific gravitational binding energy.
When the impact velocity is above the escape velocity, normal collisions are accretionary but grazing
impacts are hit and run  (little mass transfer or loss)
 (see Figure 8 by \citealt{asphaug10} and Figure 11 by \citealt{leinhardt12}).
The higher velocity impacts in Figure \ref{fig:col}b with the inner planet can cause erosion or disruption, 
and if they occur at grazing incidence then primarily erosion of surface layers is expected
 \citet{benz07,marcus10,asphaug10}.


Figures \ref{fig:exchange} and \ref{fig:76} illustrate two possible scenarios for the formation of the Kepler 36 system.
For the integration shown in  Figure \ref{fig:exchange}, the planet originally closer to the star experienced no
collisions with embryos.  
All embryo/planet collisions occurred with the outer planet.  However toward the middle of the
integration, the two planets swapped location, and the planet that had experienced collisions with 
embryos became the innermost plant.  If one or more of these collisions were sufficient high velocity and impact angle,
 the outer parts of this planet could have been stripped leaving behind the dense core that is currently Kepler 36b.
Because the other planet was protected from collisions it would have held onto its lighter elements and so would have remained at lower density.  

In the integration shown in Figure \ref{fig:76}, that ended with two planets in the 7:6 resonance,  the two planets did not exchange location.   However,  
in this integration, the inner planet experienced two collisions with embryos.   These collisions could have stripped
the inner planet of low density material.   Within the context of the planet trap,
the outer planet could continue to accrete material.  Its low density might in part be due to 
continued accretion.  

We did not allow mass to be stripped from planets during collisions in our integrations and did not take 
into account orbital debris from collisions.   Because bodies merge during our simulated collisions, in most cases
embryos were eventually incorporated into planet bodies. 
 The integrations that exchanged planet locations were from the 
X-series and so ended the integration with the more massive planet in an exterior orbit, 
opposite to the Kepler 36 system.  
Furthermore in our integrations planets only gained mass from collisions with embryos.    
A larger number of more flexible integrations could be explored to determine 
which type of scenario would best account for the Kepler 36 system origin. 

\subsection{Integrations with different mass and different number of embryos}

\begin{figure}
\includegraphics[width=3.5in,trim=0.1in 0.1in 0.1in 0.1in]{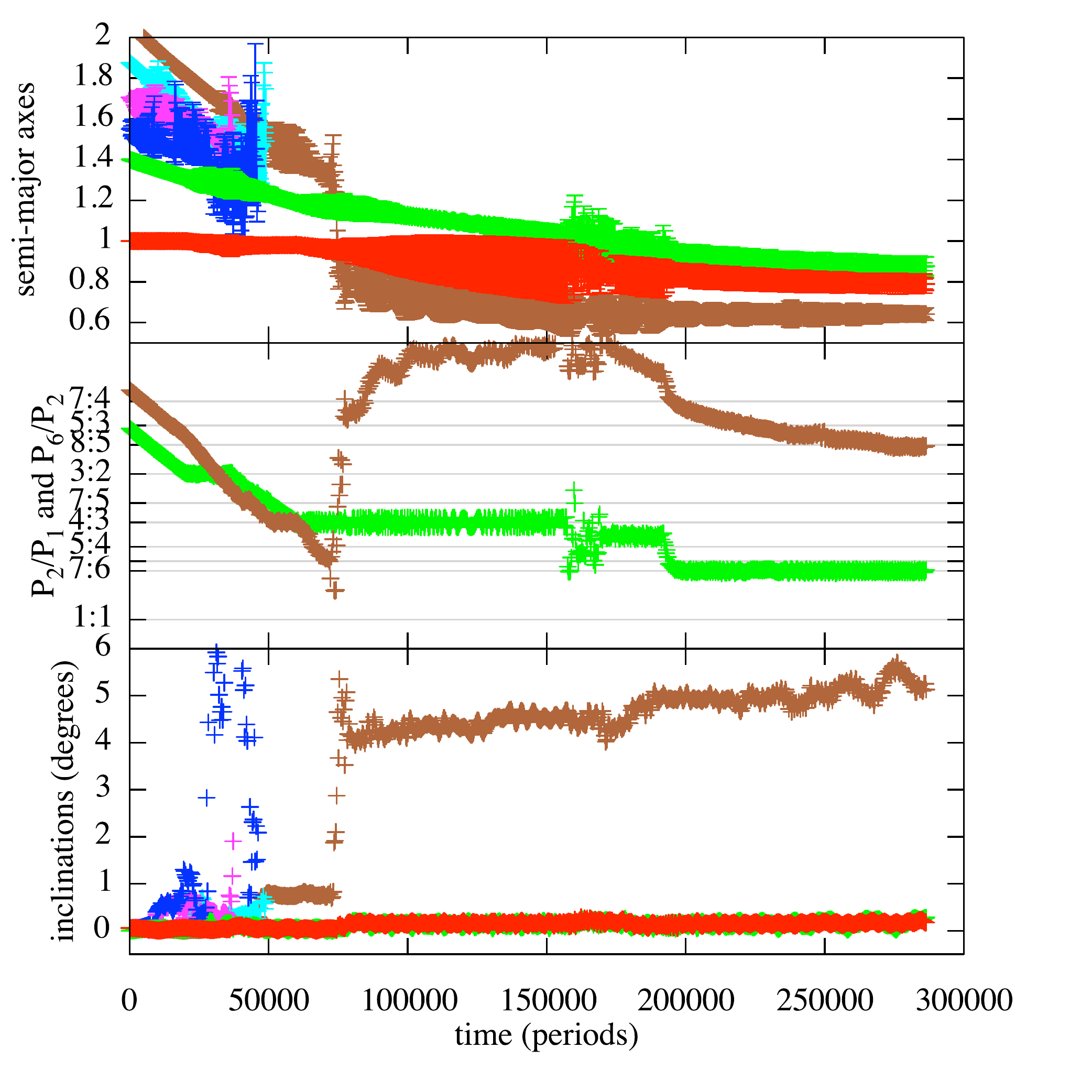}
\includegraphics[width=3.5in,trim=0.1in 0.1in 0.1in 0.1in]{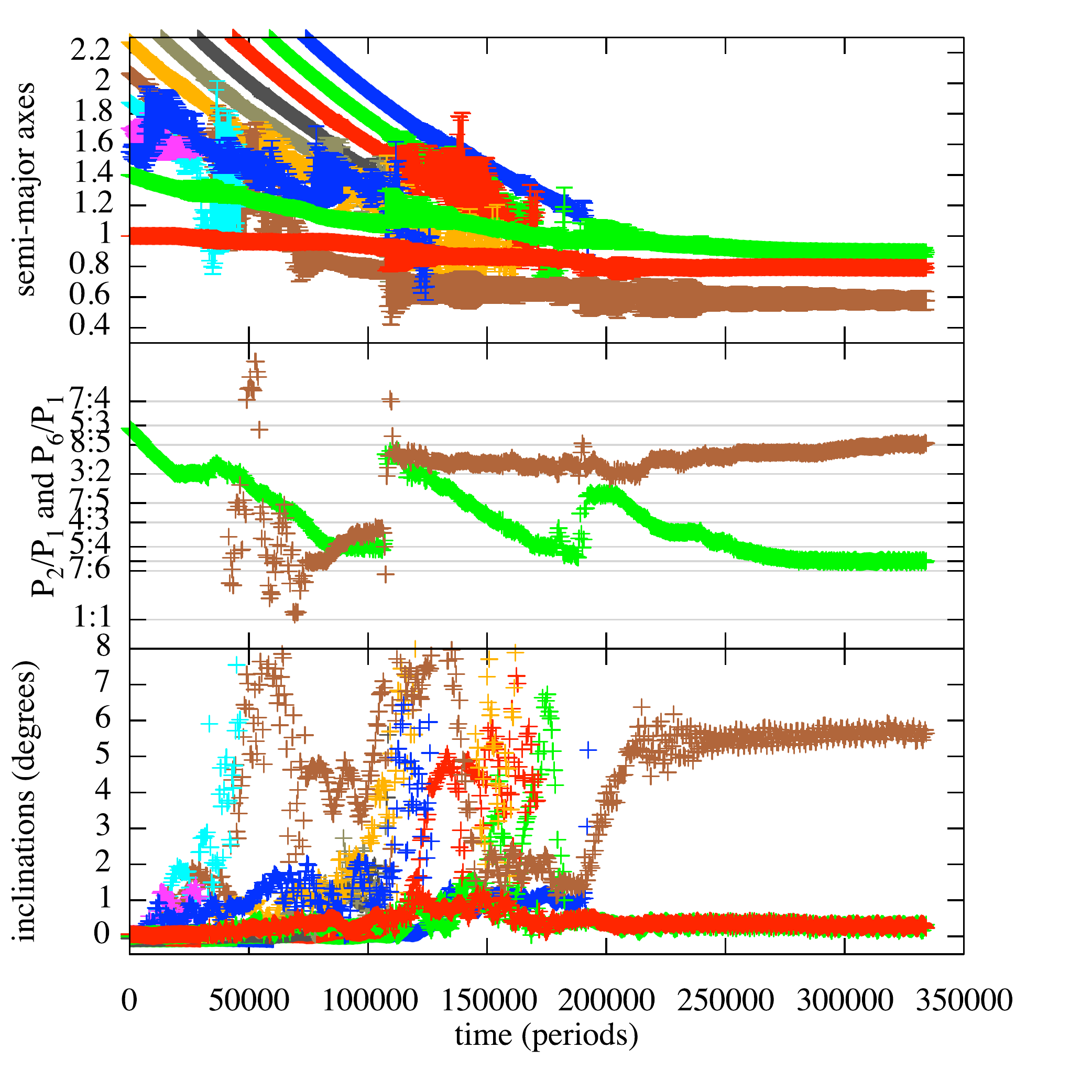}
\caption{
a) Similar to  Figure \ref{fig:exchange} except showing an integration from the A-series with 4 lower mass
embryos.
This integration illustrates that the final result can be two planets in the 7:6 resonance even with lower mass
and few embryos.
However these embryos are sufficiently low mass that collisions would not be highly disruptive.
In this simulation the inner planet did not experience any collisions.
b) Showing an integration from the Y-series with 10 embryos.
When there are many embryos, the two planets can be also scattered apart, however afterward migration
again brought them closer together and interactions with planetesimals allowed them escape strong resonances
such as the 4:3 resonance.
At the end of both of these simulations there is a remaining stable planetary embryo within the orbits of both planets.
The embryo is at a different inclination than the two planets and would not be seen in transit if the two planets were see in transit.
\label{fig:number}
}
\end{figure}

Shown in Figure \ref{fig:number} are two integrations that have fewer (4) and more (10)
embryos.  Figure \ref{fig:number}a shows an integration with lower mass embryos from the A-series
with end state of two planets in the 7:6 resonance and an internal embryo in the 8:5 resonance 
with the inner planet.  Even though there were fewer embryos, there was a time period when the two planets
interacted with two embryos and this left the two planets in nearby orbits.
The integration shown in Figure \ref{fig:number}b shows an integration from the Y-series
but with 10 embryos.   We see that twice, the planets reached the 5:4 resonance and then were
knocked apart.   However, migration then continued bringing the planets closer together.  
As embryos remained, the two planets were knocked out of strong resonances such as the 4:3 resonance,
finally reaching an end state in the 6:5 resonance with an inner stable embryo in the 8:5 resonance
with the inner planet.   The distribution of resonant outcomes does not seem to be strongly dependent on the 
number of embryos, as long as there are a few of them.  We have run integrations with larger
mass embryos and found that when $\mu_{embryo} = 1.5\times 10^{-6}$ collisions between planets
are much more frequent.

\subsection{Exiting the 7:6 resonance}

Up to this point we have explored scenarios that would place the Kepler 36 planets in the 7:6 resonance
just prior to depletion of the gaseous circumstellar disk.
Although the system's period ratio suggests the planets are  not in this resonance now, 
the proximity of the planets suggests that they might have been 
in resonance in the past.  
There is also the question of long term stability. 
Long term integrations  find that the two planet system can be
stable in the vicinity of the 7:6 resonance \citep{deck12,pardekooper13}.

Dissipation in planets from variations in the tidal force from the central star
can slowly cause orbits of planet pairs in mean motion resonance to diverge  
\citep{papa11,delisle12,lithwick12,batygin13,lee13}.
Tidal dissipation is a mechanism that could account for the many of the near resonant
Kepler planet pairs \citep{lithwick12,batygin13,lee13}.
Assuming that the Kepler 36 system was originally in 7:6 resonance, tidal forces
could have cause the period ratio to slowly increase.
We explore the tidal mechanism for causing the Kepler 36 planets to slowly diverge
from the 7:6 resonance.

The rate of eccentricity damping due to tidal dissipation in a planet can be described with a timescale
\begin{equation}
\tau_e =  {1 \over 21 \pi} {Q \over k_2} {\mu_p } \left( {  a \over R_p}\right)^5 P
\end{equation}
where $Q$ is the tidal dissipation function, and $k_2$ is a Love number (both for the planet).
The above assumes synchronous rotation and that $Q$ is independent of frequency.
Here $P$ is the rotation period of the planet. 
The above timescale is usually relevant for the innermost planet of a resonant pair as tidal
forces are a steep function of semi-major axis.  However, the timescale
is also strongly dependent on planet radius.
Using masses and radii of the Kepler 36b,c planets, we estimate
\begin{eqnarray}
\tau_{eb}  &\approx & 1.4 \times 10^8 {\rm year} {Q_b \over k_{2b}}  \nonumber \\
\tau_{ec}  &\approx & 5.8 \times 10^6 {\rm year} {Q_c \over k_{2c}}, 
\end{eqnarray}
If the two planets have similar values of $Q/k_2$, 
the tidal dissipation timescale would be much shorter for the outer more diffuse planet due to its larger radius,
than the inner rocky planet.
However, the outer planet has a low density and so could have a much higher value of $Q/k_2$ 
than a rocky or terrestrial body.

Due to tidal forces, the period ratio of the two planets is predicted to evolve as
\begin{equation}
\Delta(t) \equiv {P_2 \over P_1}  - { j \over j-1} = \left( {D_j t \over \tau_e} \right)^{1\over 3} \label{eqn:delta}
\end{equation} 
with 
\begin{equation}
D_j = {9 j^2 \over  (j-1)^3} \mu_b^2 \beta (1+\beta) C_1^2
\end{equation}
and 
\begin{equation}
\beta = {\mu_c \over \mu_b} \left({j \over j-1} \right)^{1\over 3}.
\end{equation}
We have followed the summary by \citet{lee13}, based on previous analytical studies \citep{papa11,lithwick12,delisle12,batygin13}.
The coefficient $C_1$ used by \citet{lee13} is equivalent to the function $f_{27}(\alpha)$ 
given in the appendix by \citet{M+D}.
The coefficient computed for the 7:6 resonance $C_1 \approx -5.3$.
For the Kepler 36 system we calculate $\beta = 2.1$ and $D_j \sim 5\times 10^{-8}$.
From the observed rotation periods, for the Kepler 36 planets
\begin{equation}
\Delta =  {P_c \over P_b} - {7 \over 6} =  0.006648.
\end{equation}

\citet{lee13} showed that tidal dissipation can account for the removal from resonance of Kepler
near resonant planet pairs only
if the planets are rocky and have low values of $Q/k_2$.
Following their study we invert equation \ref{eqn:delta} finding
\begin{equation}
\tau_e = {D_j t \over \Delta^3}.
\end{equation}
and use the observed value of the distance to resonance, $\Delta$, and a maximum age ($\sim 10^{10}$ years) 
for the system to place constraints on $Q/k_2$ that would allow tidal forces to account for
the distance to resonance.
For the Kepler 36 system we expect that the timescale
can be estimated using $\tau_e$ for the planet with higher dissipation rate, as
the semi-major axes and masses of the two planets are similar.
We find that a tidal damping timescale,
$\tau_e \sim {1.5 \times 10^9}$ years is required to account for the current distance from resonance.
We compare this to $\tau_{ea}$ and $\tau_{eb}$ for each planet, computed above. 
For the inner planet this implies that $Q_b/k_{2b} \lesssim 10$  and for the outer 
planet $Q_c/k_{2c} \lesssim 260$.   
The limit on the inner planet  is a factor of a few below any Solar system terrestrial planet's value and the
the limit on the outer planet a factor of 30 below any Solar system gas or ice giant.
For tidal forces to account for the distance from resonance, the planet compositions
would have to differ from those in the Solar system.

Here we have discussed only a tidal mechanism for removing the pair from resonance.
The pair could have experienced a time period of divergent migration either
due to orbit crossing debris (e.g., \citealt{moore13}) or interactions with a gas disk (e.g., \citealt{garaud07}).
Alternately the planet pair may not have been previously captured into the 7:6 resonance 
by convergent migration.


\section{Discussion and Conclusions}

In this study we have focused on origin scenarios for the Kepler 36 two planet system.
This system contains two planets of very different density, in nearby orbits, 
that are just outside the 7:6 mean motion  resonance.
We first explored a stochastic migration scenario for two planets of the same mass
as the Kepler 36b, and c planets.  Analytical estimates show
that precise adjustment of the migration rate, planet eccentricities or stochastic forcing
parameter is required so that two convergently migrating planets bypass
lower $j$ mean motion resonances, such as the 4:3 and 5:4, but capture into the 7:6
resonance.  Both simple numerical integrations and a diffusive approximation for the
stochastic forcing imply that once captured into the 7:6 resonance, the resonant system would
not be long lived. Planets soon escape resonance and are likely to collide.     
Two planets could remain in the 7:6 resonance if the turbulent disk responsible for the stochastic forcing is
 depleted soon after resonance capture.
Our findings conflict with
the recent study by \citet{pardekooper13} who have shown that long lived systems in
the 7:6 resonance can be formed through hydrodynamic simulations that induce migration, eccentricity
damping and turbulent stochastic forcing. 
Our analytical estimate for the resonant lifetime and our numerical implementation
of stochastic forcing are  based on simplistic descriptions of stochastic diffusion so
we may have underestimated the resonant lifetime.
Furthermore, the Kepler 36 system is one of a few thousand
planet candidate systems, and it is not necessary to avoid fine tuning in migration parameters or the disk depletion timescale as this type of system is not common.
 
A mechanism involving stochastic forcing 
with disk turbulence alone would not account for the density difference between the Kepler 36 planets. 
We have explored a scenario motivated by the idea of a `planet trap' \citep{masset06,morbi08}.
In this scenario two planets lie at the edge of a disk that contains a number of more rapidly migrating
planetary embryos.    We find that interactions between planetary embryos and planets can nudge the two planets
out of resonances such as the 4:3 and 5:4 resonances, leaving them in adjacent orbits.  
In our integrations with planetary embryos we have neglected
stochastic forcing due to a turbulent disk, however interactions with a gas are still assumed to take place
as we allow the embryos to migrate inwards.
Integrations with a few approximately 
Mars mass embryos display a diversity of final states, including systems with two planets in the 7:6 resonance,
two planets in the 7:5 resonance,
resonant chains, collisions between both planets and planets that have exchanged location.

We have recorded the properties of collisions that occur in the integrations.   Approximately twice as many 
planet/embryo collisions occurred with the outer planet as with the inner planet and those with the
inner planet had a wider distribution of impact velocities.  For both planets the distribution
of impact angles was wide, ranging from normal  to grazing impacts.   The distribution of impact angles
and velocities imply that the collisions can be accretionary, disruptive or strip the envelope of a planet.

Two formation scenarios for the Kepler 36 planetary system are suggested by our integrations.   
An exterior planet that is stripped by a grazing collision with an embryo,  could
leave behind a dense core that then swaps locations with a protected inner planet, 
becoming a system with a dense inner planet in orbit interior to a low density outer planet, 
like the Kepler 36 system.
Alternatively, an embryo could impact the inner planet and strip it in situ.
The `planet trap' dynamical setting could account for both the proximity of the Kepler 36 planets as well as
their large density difference.
A different type of fine tuning is likely required as the number and masses of interacting 
embryos are free parameters and the outcome is sensitive to the outcome of collisions. 
Kepler 36c's mass and radius could be consistent with an icy/heavy element core and an H/He envelope that is
0.1-0.4 times the total planet mass  
according to the 1000K models by \citet{rogers11}.
Simulations of impacts have shown that
catastrophic disruption criteria are weakly dependent on the internal composition 
of icy, rocky or strengthless colliding bodies 
\citep{leinhardt12,marcus10}.  Future studies could explore impacts with 
terrestrial mass planets containing large gaseous envelopes.  
Kepler 36's planet compositions must be considered when exploring
their collisional and accretion history.


Our integrations show that
when planets experience collisions with embryos, planet inclinations can be excited.
As transit timing variations measure the masses of more planets it may become possible to search
for correlations between planet densities and inclinations (and maybe even planet spin rates and obliquities) 
and so test the possibility that encounters
with embryos occurred in the late stages of planet formation.

Here we have explored a migration scenario involving interactions with planetary embryos to account for
both the high density contrast and near resonant location of the Kep 36 planets.
However, models of in-situ planet formation (and lacking migration) 
might account for both of these properties.
In-situ formation model predict period distributions in the 2:1 and 3:2 resonances that
are similar to the Kepler planet pair distribution at higher planet masses
\citep{petrovich13}.
\citet{owen13,lopez13} 
have recently proposed that the high density contrast of the Kepler 36 planets could be a result of
in-situ planet formation and subsequent photo-evaporation with rate that is strongly dependent on planet core mass. 

\vskip 0.3 truein

This work was in part supported by NASA grant NNX13AI27G.
We thank Matt Holman and Dan Fabrycky for brining to our attention the properties of the Kepler 36 system.
We thank Man-Hoi Lee for both hospitality and helpful comments that have improved this work.
We thank James Jenkins for correspondence on observed planet pairs in the 7:5 resonance.

\end{document}